\documentstyle[12pt,aps]{revtex}

\begin{document}

\def\fiverm{\fontfamily{\rmdefault}\fontseries{m}\fontshape{n}\tiny}

\title{\Large Proper time and path integral representations for the commutation function}

\author{S.P. Gavrilov\thanks{On leave from Tomsk Pedagogical
University, 634041 Tomsk, Russia; present  
e-mail: gavrilov@snfma1.if.usp.br} 
and D.M. Gitman\thanks{e-mail: 
gitman@snfma1.if.usp.br} \\ 
Instituto de F\'{\i}sica, Universidade de S\~ao Paulo\\
P.O. Box 66318, 05389-970 S\~ao Paulo, SP, Brasil}
\date{\today}
\maketitle

\begin{abstract}

On the example of the quantized spinor field, interacting with
arbitrary external electromagnetic field,  the commutation
function is studied. It is shown that a proper time representation is available 
 in any dimensions. Using it,  all the light
cone singularities of the function are found explicitly, generalizing the Fock formula in
four dimensions, and a path integral representation is constructed.

\end{abstract}

\newpage

\section{Introduction}

It is known that the solution of quantum field theory problems involves as
a rule different singular functions, e.g. commutation functions,
Green functions and so on. These functions are well
studied for free fields (see for example \cite{a1}). Problems appear when an
interaction is presented. In particular, it is important to study
singular functions in external backgrounds such as external fields and
curved spaces and in arbitrary dimensions. The latter may be important
for multidimensional version of field theories, which are considered
now in relation with the unification of all the interactions.
One ought to say that the commutation functions play an important role
in QFT with external backgrounds. In contrast with the case without
external backgrounds,  the perturbation theory, which takes into account
a background exactly, uses complicated (matrix) propagators. Such
propagators contain  as component parts, besides the causal and anti-causal Green
functions, the commutation functions as well \cite{a2,a3}. 
Here we present  proper time and
path integral representations for the commutation function and for
some related functions in external fields.  On the basis of the proper
time 
representation we study, in particular, light cone singularities of
the function in arbitrary dimensions. 
Traditionally QED is a
testing ground where new procedures and methods are worked through,
not infrequently creating new ideas and more profound understanding of
the structure of QFT. That is why we consider here the case of
QED with an arbitrary external electromagnetic field,
bearing in mind that the results can be extend to other theories and
backgrounds. 

Fock for the first time \cite{b1a} introduced an integral over the
proper time to present the regular on the light cone part of the 
commutation function $S(x,x')$  (in 3+1 dimensions) of the spinor
fields, interacting with an external electromagnetic field $A_\mu(x)$,
\begin{equation}\label{e1} 
S(x,x')=i[\psi(x),\bar{\psi}(x')]_+
\end{equation}
where $\psi(x)$ and $\bar{\psi}(x')$ are the electron-positron field
operators. This function  obeys the Dirac equation 
\begin{eqnarray}\label{e2} 
&&\left( \hat{{\cal P}}_{\nu} \gamma^{\nu} - m \right)
S(x,x')= 0,\;\;\hat{{\cal P}}_{\nu}=i\partial _{\nu}-gA_{\nu}(x),\\
&&[\gamma^{\mu},\gamma^{\nu}]_+ = 2\eta^{\mu\nu},  \; 
\; \eta^{\mu\nu}={\rm diag}(1,-1,-1,-1),\nonumber
\end{eqnarray}
and the initial condition
\begin{equation}\label{e3}
\left. S(x,x')\right |_{x_0=x'_0}=i\gamma^0 \delta
({\bf x}-{\bf x}').
\end{equation}
The commutation function 
$S(x,x')$ is at the same time the propagation function of the Dirac equation, i.e. it
connects solutions  $\psi(x)$ of the equation in two different time
instants,
\begin{equation}\label{e2a} 
\psi(x_0,{\bf x})=-i\int 
S(x,x')\gamma^0\psi(x'_0,{\bf x'})d{\bf x}'.
\end{equation}
Thus,  the Cauchy problem can be solved by means of the function.
Squaring the eq. (\ref{e2}) one gets
\begin{equation}\label{e4}
 S(x,x')=\left( \hat{{\cal P}}_{\nu} \gamma^{\nu} +
m \right)\Delta(x,x'), 
\end{equation}
where the function $\Delta(x,x')$ obey the equation
\begin{equation}\label{e10}
\left[( \hat{{\cal P}}_{\nu} \gamma^{\nu})^2 - m^2 \right]
\Delta(x,x')=0,
\end{equation}
and the initial conditions
\begin{equation}\label{e11}
\left. \Delta(x,x')\right |_{x_0=x'_0}=0,\;\;
\left. \partial_0\Delta(x,x')\right |_{x_0=x'_0}= \delta
({\bf x}-{\bf x}').
\end{equation}
Fock's solution of the equations (\ref{e10}), (\ref{e11}) reads
\begin{eqnarray}\label{e5}
&&\Delta(x,x')=\nonumber\\
&&\epsilon(x_0-x'_0)\left[\Theta\left((x-x')^2\right)
\Delta_R(x,x')+\frac{1}{2\pi}e^{ig\Lambda}
\delta\left((x-x')^2\right)\right],\\
&&\epsilon(x_0-x'_0)= \mbox{sign}(x_0-x'_0).\nonumber 
\end{eqnarray}
The function $\Lambda$ is the line integral of the potentials,
\begin{equation}\label{e7}
\Lambda=-\int_{x'}^{x}A_{\mu}(\tilde{x})d\tilde{x}^{\mu};
\end{equation}
$\Delta_R(x,x')$ is the Riemann function, which is presented by
means of a proper time integral
\begin{equation}\label{e6}
\Delta_R(x,x')=\int_{\Gamma_R}f(x,x',s)ds
\end{equation}
over the closed path $\Gamma_R$, (See Fig.1.)  which is a clockwise circle around
the point $s=0$ with a small enough radius, inside of which the function
 $f(x,x',s)$ has not any singularities besides the essential
singularity $s=0$.

\input prepictex
\input pictex
\input postpictex

\begin{figure}[t]
\font\thinlinefont=cmr5
\begingroup\makeatletter\ifx\SetFigFont\undefined%
\gdef\SetFigFont#1#2#3#4#5{%
  \reset@font\fontsize{#1}{#2pt}%
  \fontfamily{#3}\fontseries{#4}\fontshape{#5}%
  \selectfont}%
\fi\endgroup%
\mbox{\beginpicture
\setcoordinatesystem units <0.50000cm,0.50000cm>
\unitlength=0.50000cm
\linethickness=1pt
\setplotsymbol ({\makebox(0,0)[l]{\tencirc\symbol{'160}}})
\setshadesymbol ({\thinlinefont .})
\setlinear
%
% Fig ELLIPSE
%
\linethickness=1pt
\setplotsymbol ({\makebox(0,0)[l]{\tencirc\symbol{'160}}})
\ellipticalarc axes ratio  2.428:2.428  360 degrees 
	from 10.088 16.544 center at  7.660 16.544
%
% Fig POLYLINE object
%
\linethickness= 0.500pt
\setplotsymbol ({\thinlinefont .})
\putrule from  7.620 11.430 to  7.620 24.130
%
% arrow head
%
\plot  7.683 23.876  7.620 24.130  7.557 23.876 /
%
%
% Fig POLYLINE object
%
\linethickness= 0.500pt
\setplotsymbol ({\thinlinefont .})
\putrule from  2.540 16.510 to 18.415 16.510
%
% arrow head
%
\plot 18.161 16.447 18.415 16.510 18.161 16.574 /
%
%
% Fig POLYLINE object
%
\linethickness= 0.500pt
\setplotsymbol ({\thinlinefont .})
\plot  9.207 18.415  9.525 18.098 /
%
% arrow head
%
\plot  9.076 18.367  9.525 18.098  9.256 18.547 /
%
%
% Fig TEXT object
%
\put{\SetFigFont{9}{10.8}{\rmdefault}{\mddefault}{\updefault}O} [lB] at  6.985 16.669
%
% Fig TEXT object
%
\put{\SetFigFont{9}{10.8}{\rmdefault}{\mddefault}{\updefault}$\Gamma$} [lB] at 10.160 18.415
%
% Fig TEXT object
%
\put{\SetFigFont{6}{7.2}{\rmdefault}{\mddefault}{\updefault}R} [lB] at 10.319 18.256
%
% Fig TEXT object
%
\put{\SetFigFont{9}{10.8}{\rmdefault}{\mddefault}{\updefault}Re s} [lB] at 16.986 15.558
%
% Fig TEXT object
%
\put{\SetFigFont{9}{10.8}{\rmdefault}{\mddefault}{\updefault}Im s} [lB] at  5.874 23.495
\linethickness=0pt
\putrectangle corners at  2.515 24.155 and 18.440 11.405
\endpicture}

\caption[f1]{\label{f1}{}}
\end{figure}

%\begin{figure}[t]
%\centerline{\psfig{figure=grav.eps,width=12cm}}
%\caption{}
%\end{figure}

\noindent The function $f(x,x',s)$ obeys the
``Schr\"odinger equation''
\begin{equation}\label{e8}
i\frac{d}{ds}f(x,x',s)=\left[m^2-( \hat{{\cal P}}_{\nu} \gamma^{\nu})^2
\right]f(x,x',s),
\end{equation}
and the boundary condition
\begin{equation}\label{e9}
\lim_{s\rightarrow +0}f(x,x',s)=i\delta(x-x').
\end{equation}

One has to remark that the same function $f(x,x',s)$ appears  in the
Schwinger representation \cite{Schwin} for the causal Green function $S^c(x,x')$
(propagator) of the Dirac equation, 
\begin{equation}\label{e12a}
\left( \hat{{\cal P}}_{\nu} \gamma^{\nu} - m \right)
S^c(x,x')=-\delta(x-x').
\end{equation}
Namely, 
\begin{equation}\label{e12}
S^c(x,x')=\left( \hat{{\cal P}}_{\nu} \gamma^{\nu} +
m \right)\int_{0}^{\infty}f(x,x',s)ds\;,
\end{equation}
where at $s\rightarrow \infty$ one has to enter into the complex plane
$s$, so that $$\lim_{s\rightarrow\infty}f(x,x',s)=0.$$ Extension of the Schwinger
representation to the  curved space case was made by DeWitt \cite{a2} and
then, developing his technics, to the gauge theory \cite{b4a}.

The Schwinger representation for $S^c(x,x')$ and the Fock
representation for $S(x,x')$ differ essentially in sense of
possibilities of generalization. Thus, the Schwinger representation
retains its form for any space-time dimensionality $d$. Moreover, the
inverse operator $S^c$ can be easily presented via an exponent by means
of the Schwinger proper time representation (super-proper time
representation \cite{Git}), so that the path integral representations
follows \cite{Path,Git}.  At the same time the Fock representation 
 has the specific form (\ref{e5}) for $\Delta(x,x')$  in $d=3+1$. Besides, the
form  (\ref{e5}), which separates the light cone singular part from the regular
one,  does not give any leading consideration to write a
path integral for the commutation function, similar to one for the
propagator. 

Below we propose a  proper time representation for
the commutation function, which has an universal
form  in any dimensions. Using it, we find explicitly all the light
cone 
singularities of the commutation function in arbitrary dimensions,
generalizing the Fock's formula (\ref{e5}).
Moreover, such a representation allows one to
write a path integral for the commutation function. In
the conclusion we present similar representations for some other
singular functions of the Dirac equation on the basis of the results
obtained.

\section{Proper time representation for the commutation function}

Here we are going to write a proper time representation for the
function $\Delta(x,x')$ from the formula (\ref{e4}) in arbitrary space-time
dimensions $d\geq2$. To this end we need to find first the behavior of the
function $f(x,x',s)$ at $s\rightarrow 0$. We will use the equations
(\ref{e8}), (\ref{e9}) in d-dimensions, where
$$[\gamma^{\mu},\gamma^{\nu}]_+ = 2\eta^{\mu\nu},  \; 
\; \eta^{\mu\nu}={\rm diag}(\underbrace{1,-1,\ldots,-1}_{d}).$$
Similar to  Schwinger \cite{Schwin} we present $f(x,x's)$
as a matrix element of an evolution operator $U(s)$, 
\begin{eqnarray}\label{e14}
&&f(x,x',s)=i<x|U(s)|x'>,\;\\
&&U(s)=e^{-iHs},\; H=m^2-(\Pi_{\nu}\gamma^{\nu})^2,\nonumber
\end{eqnarray}
where  $| x >$ are eigenfunctions for some 
hermitian operators of coordinates
$X^\mu$, the corresponding canonically conjugate operators of momenta
are $P_\mu$, so that:
\begin{eqnarray}\label{e15}
&&X^\mu|x> = x^\mu |x>\,, \,\, 
 <x | x' > = \delta(x-x')\,, \,\, 
\int|x><x|dx = I\,,   \nonumber \\
&& \left[P_\mu,X^\nu \right]_- = - i \delta_\mu^\nu\,, \,\,
<x |P_\mu| x' > = -i\partial_\mu\delta(x-x')\,,  \nonumber \\ 
&&\Pi_\mu = -P_\mu - g A_\mu(X),\;\;
\left[\Pi_\mu,\Pi_\nu \right]_- = - igF_{\mu\nu}(X)\,, \,\,\,\nonumber\\
&&F_{\mu\nu}(X)=\partial_{\mu}A_{\nu}(X)-\partial_{\nu}A_{\mu}(X)\,.
\end{eqnarray}
The matrix element obeys the conditions 
\begin{eqnarray}\label{e17}
&&\left(i\frac{\partial}{\partial
x^{\mu}}-gA_{\mu}(x)\right)<x|U(s)|x'>=<x|\Pi_{\mu}U(s)|x'>,\;\nonumber\\
&&\left(-i\frac{\partial}{\partial x'^{\mu}}-gA_{\mu}(x')\right)<x|U(s)|x'>=
<x|U(s)\Pi_{\mu}|x'>.
\end{eqnarray}
Introducing the operators
$$X^{\mu}(s)=U^{-1}(s)X^{\mu}U(s),\;\gamma^{\mu}(s)=U^{-1}(s)\gamma^{\mu}U(s),\;
\Pi_{\mu}(s)=U^{-1}(s)\Pi_{\mu}U(s),$$
we come to the equations
\begin{eqnarray}\label{e16}
&&\frac{d}{ds}X^{\mu}(s)=i\left[H,X^{\mu}(s)\right]_-=2\Pi^{\mu}(s),\nonumber\\
&& \frac{d}{ds}\gamma^{\mu}(s)=i\left[H,\gamma^{\mu}(s)\right]_-\;,\;\;
\frac{d}{ds}\Pi_{\mu}(s)=i\left[H,\Pi_{\mu}(s)\right]_-\;.
\end{eqnarray}
Using the decomposition
 in
powers of $s$,  one can get for $H$: 
$$H=-\frac{1}{4s}\left[X(s)X(s)-2X(s)X(0)-X(0)X(0)\right]-i\frac{d}{2s}+O(1).$$
Then the solution  of the equations (\ref{e8}),
(\ref{e9}), (\ref{e17}) has a form 
\begin{eqnarray}\label{e18}
&&\left.f(x,x',s)\right|_{s\rightarrow 0}=f_0(x,x',s)[1+O(s)],\nonumber\\
&&f_0(x,x',s)=\frac{1}{(4\pi s)^{d/2}}\exp\left\{-i\frac{\pi}{4}(d-4)+
ig\Lambda-\frac{i}{4s}(x-x')^2\right\},
\end{eqnarray}
where  $\Lambda$ is the
$d$-dimensional line integral (\ref{e7}). Thus, one can conclude that
$f(x,x',s)$ has no singularities  
 in a small enough neighbourhood of
the point $s=0$ (excluding this point).
Based on this one can make a key observation in 3+1
dimensions. Namely, let us consider the Fock representation
(\ref{e4}), (\ref{e5}) beyound the light cone $(x-x')^2\neq 0$. In
this case the relation holds
\begin{equation}\label{e19a}
\Theta\left((x-x')^2\right)\int_{\Gamma_R}f(x,x',s)ds=
\int_{\Gamma}f(x,x',s)ds,
\end{equation}
where  $\Gamma$ (See Fig.2.) is a contour, which connects the points 
 $s=+0$ 
and $s=e^{-i\pi}0$, and passes in the lower part of complex plane $s$
 in a small enough neighbourhood of
the point $s=0$  so that (\ref{e18}) is still valid inside the contour
 $\Gamma$.

\begin{figure}[t]
\font\thinlinefont=cmr5
\begingroup\makeatletter\ifx\SetFigFont\undefined%
\gdef\SetFigFont#1#2#3#4#5{%
  \reset@font\fontsize{#1}{#2pt}%
  \fontfamily{#3}\fontseries{#4}\fontshape{#5}%
  \selectfont}%
\fi\endgroup%
\mbox{\beginpicture
\setcoordinatesystem units <0.50000cm,0.50000cm>
\unitlength=0.50000cm
\linethickness=1pt
\setplotsymbol ({\makebox(0,0)[l]{\tencirc\symbol{'160}}})
\setshadesymbol ({\thinlinefont .})
\setlinear
%
% Fig CIRCULAR ARC object
%
\linethickness=1pt
\setplotsymbol ({\makebox(0,0)[l]{\tencirc\symbol{'160}}})
\circulararc 176.529 degrees from  5.080 16.510 center at  7.699 16.589
%
% Fig POLYLINE object
%
\linethickness= 0.500pt
\setplotsymbol ({\thinlinefont .})
\putrule from  7.620 11.430 to  7.620 24.130
%
% arrow head
%
\plot  7.683 23.876  7.620 24.130  7.557 23.876 /
%
%
% Fig POLYLINE object
%
\linethickness= 0.500pt
\setplotsymbol ({\thinlinefont .})
\putrule from  2.540 16.510 to 18.415 16.510
%
% arrow head
%
\plot 18.161 16.447 18.415 16.510 18.161 16.574 /
%
%
% Fig POLYLINE object
%
\linethickness=1pt
\setplotsymbol ({\makebox(0,0)[l]{\tencirc\symbol{'160}}})
\putrule from  5.080 16.510 to 10.319 16.510
%
% Fig POLYLINE object
%
\linethickness= 0.500pt
\setplotsymbol ({\thinlinefont .})
\plot  9.684 14.764  9.366 14.446 /
%
% arrow head
%
\plot  9.636 14.895  9.366 14.446  9.815 14.716 /
%
%
% Fig TEXT object
%
\put{\SetFigFont{9}{10.8}{\rmdefault}{\mddefault}{\updefault}O} [lB] at  6.985 16.669
%
% Fig TEXT object
%
\put{\SetFigFont{9}{10.8}{\rmdefault}{\mddefault}{\updefault}Re s} [lB] at 16.986 15.558
%
% Fig TEXT object
%
\put{\SetFigFont{9}{10.8}{\rmdefault}{\mddefault}{\updefault}Im s} [lB] at  5.874 23.495
%
% Fig TEXT object
%
\put{\SetFigFont{9}{10.8}{\rmdefault}{\mddefault}{\updefault}$\Gamma$} [lB] at  9.842 13.970
\linethickness=0pt
\putrectangle corners at  2.515 24.155 and 18.440 11.405
\endpicture}

\caption{}
\end{figure}

%\begin{figure}[t]
%\centerline{\psfig{figure=grav2.eps,width=12cm}}
%\caption{}
%\end{figure}

\noindent Then the function $\Delta(x,x')$  can be written
in 3+1 dimensions at
$(x-x')^2\neq 0$ in the following form
\begin{equation}\label{e19}
\Delta(x,x')=\epsilon(x_0-x'_0)\int_{\Gamma}f(x,x',s)ds.
\end{equation}
It turns out that (\ref{e19}) is valid on the light cone as well and,
moreover, in any dimensions. Below we are going to prove this
statement.

First of all one can remark that (\ref{e18}) implies
\begin{equation}\label{e18a}
\lim_{s\rightarrow e^{-i\pi}0}f(x,x',s)=-i\delta(x-x').
\end{equation}
Together with (\ref{e8}), (\ref{e9}) and (\ref{e18a}) this allows one
to verify that the expression (\ref{e19}) obeys the equation
(\ref{e10}) at $x_0\neq x'_0.$

Now we have to study the behavior of the function (\ref{e19}) at
$x_0\rightarrow x'_0.$
Using the representation (\ref{e18}) let us select  all
the light cone singularities in (\ref{e19}),
\begin{eqnarray}
&&\Delta(x,x')=I_R(x,x')+\sum_{n=0}^{[d/2]-1}I^{(n)}(x,x'),\label{e20}\\
&&I_R(x,x')=\epsilon(x_0-x'_0)\int_{\Gamma}f_R(x,x',s)ds,\label{e21}\\
&&f_R(x,x',s)=f(x,x',s)-\sum_{n=0}^{[d/2]-1}f^{(n)}(x,x',s),\nonumber\\
&&I^{(n)}(x,x')=\epsilon(x_0-x'_0)\int_{\Gamma}f^{(n)}(x,x',s)ds,\label{e22}\\
&&f^{(n)}(x,x',s)=\frac{1}{n!}\frac{d^n}{ds^n}\left[\frac{f(x,x',s)}{f_0(x,x',s)}
\right]_{s=0}s^nf_0(x,x',s).\nonumber
\end{eqnarray}
Here $I_R(x,x')$ is a regular on the light cone function, which is
zero at $(x-x')^2\leq 0.$  All the singularities are concentrated in
the functions $I^{(n)}(x,x')$. It is convenient to make a change of
variables $t=s^{-1}$ in the expression  (\ref{e22}) to present the
latter in the following form 
\begin{eqnarray}\label{e23}
&&I^{(n)}(x,x')=\epsilon(x_0-x'_0)a^{(n)}(x,x')\left(\frac{d}{d\tau}\right)^{[\frac{d-2}{2}]-n}
\left.Y\left(\tau-({\bf x}-{\bf x}')^2\right)\right|_{\tau=(x_0-x'_0)^2},\nonumber\\
&&a^{(n)}(x,x')=\pi^{-[\frac{d-1}{2}]}e^{ig\Lambda} (4i)^{-n} 
\frac{1}{4n!}\frac{d^n}{ds^n}\left[\frac{f(x,x',s)}{f_0(x,x',s)}
\right]_{s=0}\;,
\end{eqnarray}
where the function $Y\left(\tau-({\bf x}-{\bf x}')^2\right)$ has
different forms for even  and for odd $d$. Namely, for even $d$,
\begin{equation}\label{e24}
Y\left(\tau-({\bf x}-{\bf x}')^2\right)=Y_{even}\left(\tau-({\bf x}-{\bf
x}')^2\right)=
2\Theta\left(\tau-({\bf x}-{\bf x}')^2\right)
0,\end{equation}
and for odd $d$,
\begin{eqnarray}\label{e25}
&&Y\left(\tau-({\bf x}-{\bf x}')^2\right)=
Y_{odd}\left(\tau-({\bf x}-{\bf x}')^2\right)\nonumber\\
&&=e^{i\frac{\pi}{4}}\int_{e^{i\pi}\infty}^{\infty} (4\pi
t)^{\frac{1}{2}}\exp\left\{-\frac{i}{4\pi}\left[\tau-({\bf x-x'})^2\right]t\right\}dt.
\end{eqnarray}
Now one can see that for even $d$ the function
$I^{(\frac{d}{2}-1)}(x,x')$ can be expressed via the $\Theta$-function
(\ref{e24}), whereas the rest functions $I^{(n)}(x,x')$ are
concentrated on the light cone. Thus, the function (\ref{e20}) is zero
for points, which can not be causally connected, i.e. for  $(x-x')^2< 0.$

Consider the contribution of the distributions  (\ref{e24}) and
(\ref{e25}) to an integral with some continuous functions on the
coordinates $({\bf x-x'})$ in the case $(x_0-x'_0)\rightarrow 0.$ In this
case the distributions are zero beyond the sphere of the radius $|x_0-x'_0|+0$,
thus the formula (\ref{e24}) can be written in the form
\begin{equation}\label{e26}
Y_{even}\left(\tau-({\bf x}-{\bf x}')^2\right)=2V(\sqrt{\tau})\delta({\bf x-x'}),
\end{equation}
where $V(r)$ is the volume of the $d-1$ sphere with the radius $r$,
\begin{equation}\label{e27}
V(r)=cr^{d-1},\;\;c=\pi^{\frac{d-1}{2}}\Gamma^{-1}\left(\frac{d+1}{2}\right),
\end{equation}
and $\Gamma(x)$ is the gamma-function. The least power of 
$(x_0-x'_0)$ in the expression (\ref{e20}) comes from 
$I^{(0)}(x,x')$, the latter can be derived from (\ref{e23}),
(\ref{e26})  and has the form 
$I^{(0)}(x,x')=e^{ig\Lambda}(x_0-x'_0)\delta({\bf x-x'}).$ In case $d\geq 4$, the
next power of   $(x_0-x'_0)$ comes from the function
$I^{(1)}(x,x')\sim e^{ig\Lambda}(x_0-x'_0)^3\delta({\bf x-x'}).$ At $d=2$ the same
power of $(x_0-x'_0)$ comes from the function $I_R(x,x')$ defined by eq.
(\ref{e21}). Thus, we can write for any even $d$ at
$(x_0-x'_0)\rightarrow 0$:
\begin{equation}\label{e28}
\left. \Delta(x,x')\right |_{x_0\rightarrow x'_
0}=e^{ig\Lambda}\left[x_0-x'_0+O\left(x_0-x'_0)^3\right)\right]\delta({\bf x-x'}).
\end{equation}

One can see that (\ref{e28}) is a continuous function of the time
$(x_0-x'_0)$ together with its first derivatives, and obeys the equation
(\ref{e10})  at $(x_0-x'_0)\rightarrow 0$, and the initial conditions
(\ref{e11}).

The expression (\ref{e25}) at $(x_0-x'_0)\rightarrow 0$ ($d$ is odd) can
be presented in the form
\begin{equation}\label{e29}
Y_{odd}\left(\tau-({\bf x}-{\bf x}')^2\right)=B(\tau)\delta({\bf x-x'}),
\end{equation}
where
\begin{equation}\label{e30}
B(\tau)=\int_{V(r_0)}Y_{odd}(\tau-{\bf y}^2)d{\bf y}.
\end{equation}
The integration in (\ref{e30}) is going over the volume of $d-1$
sphere with the radius $r_0=|x_0-x'_0|+0$. The integral (\ref{e30}) is
reducing to one over the radius $r$ only,
\begin{equation}\label{e31}
B(\tau)=(d-1)c\int_{0}^{r_0}Y_{odd}(\tau-r^2)r^{d-2}dr,
\end{equation}
where $c$ was defined in (\ref{e27}) and we remember that $r_0>\tau$.
 The former can be calculated and
presented in the form
\begin{equation}\label{e32}
B(\tau)=e^{i\frac{\pi d}{4}}\pi^{\frac{d-2}{2}}
\int_{e^{i\pi}\infty}^{\infty}t^{-\frac{d}{2}}e^{-i\tau t}dt.
\end{equation}

Similar to the even case the least power of $(x_0-x'_0)$ in (\ref{e20})
comes from the function $I^{(0)}(x,x')$. By means of (\ref{e29}) and
(\ref{e32}) the latter can be written as
\begin{equation}\label{e33}
I^{(0)}(x,x')=\epsilon(x_0-x'_0)\delta({\bf x-x'})
\frac{e^{i\frac{3\pi}{4}}e^{ig\Lambda}}{4\sqrt{\pi}}
\int_{e^{i\pi}\infty}^{\infty}t^{-\frac{3}{2}}e^{-i(x_0-x'_0)^2 t}dt.
\end{equation}
Using the representation
$$t^{-\frac{1}{2}}=e^{-i\frac{\pi}{4}}\frac{2}{\sqrt{\pi}}\left(\int_{0}^{\sqrt{(x_0-x'_0)^2-0}}
e^{iz^2t}dz+
\int_{\sqrt{(x_0-x'_0)^2+0}}^{\infty}e^{iz^2t}dz\right)$$
in (\ref{e33}) and changing the order of the integration over $t$ and
$z$, we get finally: $I^{(0)}(x,x')=e^{ig\Lambda}(x_0-x'_0)\delta({\bf x-x'}).$ In
the same manner one can verify that the next power of $(x_0-x'_0)$ at
 $d\geq 5$   comes from the function
$I^{(1)}(x,x')\sim e^{ig\Lambda}(x_0-x'_0)^3\delta({\bf x-x'}).$  In the case $d=3$ the same
power of $(x_0-x'_0)$ comes from the function 
(\ref{e21}). That is why the same dependence (\ref{e28}) holds at any
odd $d$.

Thus, we have shown that the function (\ref{e19}) obeys the equation
(\ref{e10}) and the initial conditions (\ref{e11}) in any dimensions
$d$. Then the commutation function can be written in an universal
form in any dimensions (by means (\ref{e4}), (\ref{e19})),
\begin{equation}\label{e19b}
 S(x,x')=\epsilon(x_0-x'_0)\left( \hat{{\cal P}}_{\nu} \gamma^{\nu} +
m \right)\int_{\Gamma}f(x,x',s)ds. 
\end{equation}
Here we have used the initial conditions (\ref{e11}) to put
$\epsilon(x_0-x'_0)$
before the operator $ \hat{{\cal P}}_{\nu} \gamma^{\nu} + m. $

It was already seen from (\ref{e20}) - (\ref{e25}) that the representation
(\ref{e19}) is convenient to select the light cone singularities. In
case of even $d$ one can also get  $d$-dimensional generalization
of the Fock representation. To this end let us write the function
$I_R(x,x')$ from (\ref{e21}) by means of an integral over the closed
path $\Gamma_R$, defined in (\ref{e6}),
\begin{equation}\label{e34}
I_R(x,x')=\epsilon(x_0-x'_0)\Theta\left((x-x')^2\right)\int_{\Gamma_R}f_R(x,x',s)ds.
\end{equation}
From $f_R(x,x',s)$ only the term $f^{(n)}(x,x',s)$ with
$n=\frac{d}{2}-1$ gives nonzero contribution, namely, 
$$\Theta\left((x-x')^2\right)\int_{\Gamma_R}f^{(\frac{d}{2}-1)}(x,x',s)ds=
\int_{\Gamma}f^{(\frac{d}{2}-1)}(x,x',s)ds.$$
That allows one to rewrite (\ref{e20}) for even $d$ in the form
\begin{eqnarray}\label{e35}
&&\Delta(x,x')=\epsilon(x_0-x'_0)\left[\Theta\left((x-x')^2\right)
\Delta_R(x,x')\right.\nonumber\\
&&+\left.\sum_{n=0}^{(d-4)/2}
2a^{(n)}(x,x')\left.\left(\frac{d}{d\tau}\right)^{\frac{d-4}{2}-n}
\delta\left(\tau-({\bf x-x'})^2\right)\right|_{\tau=(x_0-x'_0)^2}\right],
\end{eqnarray}
where $a^{(n)}(x,x')$ are defined in (\ref{e23}), and $\Delta_R(x,x')$
is $d$-dimensional Riemann function, defined by the integral
(\ref{e6}). 
At $d=4$ this expression coincides with the Fock's one
(\ref{e5}). 

The expression for the commutation function
$i\left[\phi(x),\phi^{\dagger}(x')\right]_-$ 
of the scalar fields $\phi(x)$ and $\phi^{\dagger}(x')$
one can derive from the representation
(\ref{e19}) for the function $\Delta(x,x')$, putting formally all the
$\gamma$-matrices to zero.
We do not also see any difficulties to extend the results obtained to the
curved space and gauge theories using the Schwinger-DeWitt technics
\cite{a2,b4a}. 

\section{Path integral representation for the commutation function}

Here we are going to discuss a path integral representation for
the commutation function at $d=4$.  For our purpose, it is 
convenient  to deal with the transformed by
$\gamma^5=\gamma^0\gamma^1
\gamma^2\gamma^3$ function 
$\tilde{S}(x,x') = S(x,x')\gamma^5\,\,\,$, which obeys the 
properly transformed Dirac equation 
\begin{equation}\label{d1} 
\left( \hat{{\cal P}}_\nu \tilde{\gamma}^\nu - m\gamma^5 \right)
\tilde{S}(x,x')= 0,
\end{equation}
and the initial condition
\begin{equation}\label{d1a}
\left. \tilde{S}(x,x')\right |_{x_0=x'_0}=-i\tilde{\gamma}^0 \delta
({\bf x}-{\bf x}'),
\end{equation}
where $\hat{{\cal P}}_{\nu}=i\partial _{\nu}-gA_{\nu}(x),$ and $\tilde{\gamma}^
\nu = \gamma^5\gamma^\nu.$ The matrices $\tilde{\gamma}^\nu$  have the same 
commutation relations as initial ones $\gamma^\nu ,\; \;
\left[\tilde{\gamma}^\mu,\tilde{\gamma}^\nu\right]_+=2\eta^{\mu\nu}$.
 For all the $\gamma$-matrices ($\tilde{\gamma}^5=\gamma^5$) we
have $[\tilde{\gamma}^m,\tilde{\gamma}^n]_+ = 2\eta^{mn}, \;\; m,n=\overline{0,3},5; \; 
\; \eta^{mn}={\rm diag}(1,-1,-1,-1,-1)$.

If one presents the function $ \tilde{S}(x,y)$ in the  form
\begin{equation}\label{d1b}
 \tilde{S}(x,x')=-\left( \hat{{\cal P}}_\nu \tilde{\gamma}^\nu -
m\gamma^5 \right)\tilde{\Delta}(x,x'),
\end{equation}
then the function $\tilde {\Delta }$ obeys the equation
\begin{equation}\label{d1c}
\left( \hat{{\cal P}}_\nu \tilde{\gamma}^\nu - m\gamma^5 \right)^2
\tilde{\Delta}(x,x')=\left[ (\hat{{\cal P}}_\nu \gamma ^\nu)^2 - 
m^2 \right] \tilde{\Delta}(x,x')=0.
\end{equation}
One can  remark that according to the definition and to the eq. (\ref{e10}),(\ref{e11}) there is
a relation $\tilde{\Delta}(x,x')=-\gamma^5 \Delta (x,x') \gamma^5$, which
allows on  to conclude that the functions $\tilde{\Delta}(x,x')$ and $
\Delta (x,x')$ obey the same initial conditions. Because  they obey
also the same equation they coincide. Thus, one can write, using the
results obtained before (\ref{e19b}), 
\begin{equation}\label{d1d}
 \tilde{S}(x,x')=-\epsilon(x_0-x_0')\int_{\Gamma}\left( \hat{{\cal P}}_\nu \tilde{\gamma}^\nu -
m\gamma^5 \right)f(x,x',s)ds \;.
\end{equation}
By means of the representation (\ref{e14}) for the function $f(x,x',s)$, where one
can replace the operator $\left[ (\Pi_\nu \gamma ^\nu)^2 - 
m^2 \right]$ by one $\left( \Pi_\nu \tilde{\gamma}^\nu -
m\gamma^5 \right)^2$, and introducing the operator $\left( \hat{{\cal
P}}_\nu \tilde{\gamma}^\nu -m\gamma^5 \right)$ under the sign of  the matrix element,
we get 
\begin{eqnarray}\label{d1e}
&&\tilde{S}(x,x')=-i\;\epsilon(x_0-x_0')\int_{\Gamma}<x|\left(
\Pi_{\nu} 
\tilde{\gamma}^\nu -
m\gamma^5 \right)\nonumber\\
&&\exp\left\{i\left( \Pi_\nu \tilde{\gamma}^\nu - m\gamma^5 \right)^2
\right\}|x'>ds \;.
\end{eqnarray}
The operator $\left( \Pi_\nu \tilde{\gamma}^\nu -
m\gamma^5 \right)$ can be presented via a Grassmannian integral, 
\[
\left( \Pi_\nu \tilde{\gamma}^\nu -
m\gamma^5 \right)=i \int  e^{i\chi \left( \Pi_\nu \tilde{\gamma}^\nu -
m\gamma^5 \right)} d\,\chi\,,  
\]
where $\chi$ is a Grassmann
variable, which  anticommutes with $\gamma$ matrices by
the definition. Here and in what follow integrals 
over Grassmann variables are understood as Berezin's integrals
\cite{Grass}. Thus, the commutation function (\ref{d1e}) takes the form  
\begin{equation}\label{d5}
\tilde{S}=\tilde{S}(x_{out},x_{in}) =\epsilon(x_{out}^0,x_{in}^0)
\int_{\Gamma} \, ds 
\int \langle x_{\rm out} | e^{-i\hat{\cal H}(s,\chi)}|x_{\rm in} 
\rangle d\chi\,\,,
\end{equation}
where
\begin{equation}\label{d5a}
\hat{{\cal H}}(s,\chi)=s \left(
m^2 - \Pi^2 +\frac{ig}{2}F_{\alpha\beta} \tilde{ \gamma}^\alpha
\tilde{\gamma}^\beta\right) + 
\left(\Pi_\nu\tilde{\gamma}^\nu - m \tilde{\gamma}^5 \right)\chi\; .
\end{equation}

Now one can present the matrix element entering in the expression
(\ref{d5}) by means of a path integral. First, we write 
$\exp -i\hat{{\cal H}} = \left(\exp -i\hat{{\cal H}}/N \right)^N, $ 
and then insert $(N-1)$
resolutions of identity $\int|x><x|dx = I$ between all the 
operators $\exp -i\hat{{\cal H}}/N$. Besides, we introduce
$N$ additional integrations over $s$ and $\chi$ to transform then the
ordinary integrals over these variables into the corresponding path-integrals,
\begin{eqnarray}\label{d6}
&& \tilde{S} = \epsilon(x_{out}^0,x_{in}^0) \lim_{N\rightarrow \infty}\int_{\Gamma}d\, s_0 
\int d\, \chi_0 d\, x_1 ... d\, x_{N-1}
d\, s_1 ... d\, s_N d\, \chi_1 ... d\, \chi_N \nonumber \\
&&\times \prod_{k=1}^{N}\langle x_k |
e^{-i\hat{{\cal H}}(s_k,\chi_k)\Delta \tau} | x_{k-1} \rangle 
\delta(s_k-s_{k-1})\delta(\chi_k-\chi_{k-1})\; ,
\end{eqnarray}
where $\Delta \tau = 1/N$, $x_0=x_{\rm in}$, $x_N=x_{\rm out}$. Bearing in 
mind the limiting process, one can calculate the matrix elements 
from (\ref{d6}) approximately,
\begin{equation}\label{d7}
\langle x_k |
e^{-i\hat{{\cal H}}(s_k,\chi_k)\Delta \tau} | x_{k-1} \rangle \approx
\langle x_k |
1{-i\hat{{\cal H}}(s_k,\chi_k)\Delta \tau} | x_{k-1} \rangle,
\end{equation}
using the resolution of identity $\int|p><p| d\,p$, where
\[
P_\mu|p> = p_\mu |p>, \, 
<p | p' > = \delta^4(p-p'),\;
<x | p > = \frac{1}{(2\pi)^2}e^{ipx}. 
\]
In this 
connection it is  important to notice that the operator
$\hat{{\cal H}}(s_k,\chi_k)$ has originally the symmetric form in the 
operators $\hat{x}$ and $\hat{p}$. Indeed, the only one term in 
$\hat{{\cal H}}(s_k,\chi_k)$, which contains products of these operators
is $[P_\alpha,A^\alpha(X)]_+$. One can verify that this is
maximal symmetrized expression, which can be combined from entering operators
(see remark in  \cite{ord1}). Thus, one can write
$\hat{{\cal H}}(s,\chi) = {\rm Sym}_{(\hat{x},\hat{p})}\,\, 
{\cal H}(s,\chi,\hat{x},\hat{p})$,
where ${\cal H}(s,\chi,{x},{p})$ is the Weyl symbol of the 
operator $\hat{{\cal H}}(s,\chi)$ in the sector of coordinates and 
momenta, ${\cal H}(s,\chi,x,p) =  s \left(m^2-{\cal P}^2 +
\frac{ig}{2}F_{\alpha\beta}  \tilde{\gamma}^\alpha\tilde{\gamma}^\beta\right)
+\left(  {\cal P}_\nu\tilde{\gamma}^\nu - m\gamma^5 \right)\chi\,,$
and ${\cal P}_\nu= -p_\nu-gA_\nu(x)\,$.
That is a general statement \cite{ord2}, which can be easily checked in that
concrete case by direct calculations, that the matrix elements (\ref{d7})
are expressed in terms of the Weyl symbols in the middle point 
$\overline{x}_k = (x_k+x_{k-1})/2$.
Taking all that into account, one can see that in the limiting process the 
matrix elements (\ref{d7}) can be replaced by the expressions 
\begin{equation}\label{d8}
\int \frac{d\,p_k}{(2\pi)^4}\exp i \left[
p_k\frac{x_k-x_{k-1}}{\Delta\tau} - {\cal H}(s_k,\chi_k,
\overline{x}_k,p_k) \right]\Delta\tau\,,
\end{equation}
which are non-commutative  due to the $\gamma$-matrix structure and are
situated in (\ref{d6}) so that the numbers $k$ increase 
from the right to the left. For the
two $\delta$-functions, accompanying each matrix element (\ref{d7}) in the
expression (\ref{d6}), we use the integral representations
$$
\delta(s_k-s_{k-1})\delta(\chi_k-\chi_{k-1}) = 
\frac{i}{2\pi} \int e^{i\left[ 
\pi_k\left(s_k-s_{k-1}\right)+
\nu_k\left(\chi_k   -\chi_{k-1}   \right)
\right]}d\, \pi_k d\, \nu_k,
$$
where $\nu_k$ are odd variables. Then we attribute formally to 
$\gamma$-matrices, entering into  (\ref{d8}), index $k$, and
then we attribute to all quantities the ``time'' $\tau_k$, according the
index $k$ they have, $\tau_k=k\Delta\tau $, so that $\tau \in [0,1]$.
Introducing the T-product, which acts on $\gamma$-matrices, it is possible 
to gather all the expressions, entering in  (\ref{d6}), 
in one exponent and deal then with the $\gamma$-matrices
like with odd  variables. Thus, we get for the right side of (\ref{d6})
\begin{eqnarray}\label{d9}
&&\tilde{S} =\epsilon(x_{out}^0,x_{in}^0) {\rm T}\int_{\Gamma} \, ds_0 \int  
d\chi_{0}\int_{x_{in}}^{x_{out}}Dx \int Dp \int_{s_0}Ds
\int_{\chi_0}D\chi\int D\pi \int D\nu \nonumber\\
&&\times\exp \left\{i\int_0^1 \left[ s \left({\cal P}^2 - m^2 
-\frac{ig}{2}F_{\alpha\beta}  \tilde{\gamma}^\alpha\tilde{\gamma}^\beta
\right)\right.\right.\nonumber\\
&&+ \left.\left.
\left( m\gamma^5 - {\cal P}_\nu\tilde{\gamma}^\nu \right)\chi + 
p\dot{x} + \pi\dot{s} + \nu\dot{\chi}\right]d\tau \right \} \,,  
\end{eqnarray}
where   $x$,  $p$,  $s$,  $\pi$ ,
are even and $\chi, \;\nu$ are odd trajectories, obeying the boundary conditions
 $\,\,\,x(0)=x_{\rm in}$, $\,\,\,x(1)=x_{\rm out}$, 
$\,\,\,s (0) = s_0$, 
$\,\,\,\chi(0) = \chi_0$. The operation of T-ordering  
acts on the $\gamma$-matrices, which suppose formally to
depend on time $\tau$. The expression (\ref{d9}) can be reduced to:
\begin{eqnarray*}
&& \tilde{S} =\epsilon(x_{out}^0,x_{in}^0) \int_{\Gamma} \, ds_0
\int d\chi_{0}\int_{x_{in}}^{x_{out}}Dx \int Dp \int_{s_0}Ds
\int_{\chi_0}D\chi\int D\pi \int D\nu\\
&& \exp \left\{i\int_0^1 \left[
s\left( {\cal P}^2 -m^2 \right.\right. \right. 
\left.\left.\left.-\frac{ig}{2}F_{\alpha\beta} \frac{\delta_l}{\delta 
\rho_\alpha}\frac{\delta_l}{\delta \rho_\beta}\right)  \right.\right.\\
&&+\left.\left.\left( m\frac{\delta_l}{\delta \rho_5} 
   - {\cal P}_\nu\frac{\delta_l}{\delta \rho_\nu}\right)\chi +\left. 
p\dot{x} + \pi\dot{s} + \nu\dot{\chi}\right]d\tau\right\} 
{\rm T}\exp \int_0^1\rho_n(\tau)\tilde{\gamma}^n d\tau \right|_{\rho=0},
\end{eqnarray*}
where five odd sources $\rho_n(\tau)$ are introduced, which  
anticommute with the $\gamma$-matrices by definition. One can present the 
quantity ${\rm T}\exp \int_0^1 \rho_n(\tau)\tilde{\gamma}^n 
d\tau$  via a  path integral over odd trajectories \cite{Git}, 
\begin{eqnarray}\label{d10}
&&{\rm T}\exp \int_0^1\rho_n(\tau)\tilde{\gamma}^n d\tau  = \exp\left(i\tilde{\gamma}^n
\frac{\partial_l}{\partial\theta^n}   \right)
\int_{\psi(0)+\psi(1)=\theta}\exp \left[ \int_0^1 \left( 
\psi_n\dot{\psi}^n \right.\right.\nonumber\\
&&-\left.\left. 2i\rho_n\psi^n\right) d\tau \right. 
+ \left.\left.\psi_n(1)\psi^n(0)\right]{\cal D}\psi\right|_{\theta=0},\;\;\\
&&{\cal D}\psi=D\psi\left[\int_{\psi (0)+\psi (1)=0}D\psi \exp\left\{\int^{1}
_{0}\psi_{n}\dot{\psi}^{n}d\tau\right\}\right]^{-1} \; ,\nonumber
\end{eqnarray}
where $\theta^n$ are odd variables, anticommuting with $\gamma$-matrices, and 
$\psi^{n}(\tau)$ are odd trajectories of integration, obeying the boundary 
conditions, which are pointed out below the signs of integration. 
Using (\ref{d10}) we get the Hamiltonian path integral 
representation for the commutation function:
\begin{eqnarray*}\label{path}
&&\tilde{S} =\epsilon(x_{out}^0,x_{in}^0)\exp\left(i\tilde{\gamma}^n
\frac{\partial_l}{\partial\theta^n} \right)\int_{\Gamma}  ds_0 \int
d\chi_{0} \int_{s_{0}}Ds
\int_{\chi_{0}}D\chi \int_{x_{in}}^{x_{out}}Dx \int Dp \\
&&\int D\pi \int
D\nu  \times \int_{\psi(0)+\psi(1)=\theta} {\cal D}\psi \exp \left\{i\int_0^1 
\left[ s\left({\cal P}^2 - m^2
+2igeF_{\alpha\beta}\psi^\alpha\psi^\beta\right)\right.\right.\\
&&+ \left.\left.
2i\left({\cal P}_\alpha
\psi^\alpha - m\psi^5\right)\chi \right.\right.   \left.\left.
-i\psi_n\dot{\psi}^n +p\dot{x} + 
\pi \dot{s} +\nu \dot{\chi}
\right] d\tau +\left.\psi_n(1)\psi^n(0) \right\}\right|_{\theta=0}, 
\end{eqnarray*}
Integrating over momenta in the path integral, we get 
\begin{eqnarray}
&&\tilde{S} =\epsilon(x_{out}^0,x_{in}^0)\exp\left(i\tilde{\gamma}^n
\frac{\partial_l}{\partial\theta^n} \right)\int_{\Gamma}  de_0 \int
d\chi_{0}G(e_0,\chi_0,x_{out}^0,x_{in}^0)\;,\label{d11} \\
&&G(e_0,\chi_0,x_{out}^0,x_{in}^0)=\int_{e_{0}}
De\int_{\chi_{0}}D\chi \int_{x_{in}}^{x_{out}}Dx \int D\pi \int 
D\nu \int_{\psi(0)+\psi(1)=\theta} {\cal D}\psi \nonumber \\  
&&\times M(e) \exp\left\{i\int_{0}^{1}\left[-\frac{\dot{x}^{2}}
{2e}-\frac{e}{2}m^{2}-g\dot{x}A(x)+iegF_{\mu
\nu}(x)\psi^{\mu}\psi^{\nu} 
\right.\right.\nonumber\\
&&+\left.\left.
i\left(\frac{\dot{x}_{\mu}\psi^{\mu}}{e}-m\psi^{5}\right)\chi\right.\right. 
\left.\left.-i\psi_{n}\dot{\psi}^{n}+\pi \dot{e}+\nu \dot{\chi}\right]d\tau
+ \left.\psi_{n}(1)\psi^{n}(0)\right\}\right|_{\theta=0}\;,\label{d11a}
\end{eqnarray}
\noindent where $M(e)$ is the integration measure, 
\begin{equation}\label{d12}
M(e)=\int Dp\exp\left\{ \frac{i}{2}\int_{0}^{1} ep^{2}d\tau\right\} \;.
\end{equation}
The exponent in the integrand (\ref{d11a}) can be considered as an 
effective and non-degenerate Lagrangian action of a spinning particle in an 
external field. It consists of two principal parts. The first one,
which unites two summand with the derivatives of $e$ and $\chi$, can be 
treated as a gauge fixing term and corresponds to the gauge 
conditions $\dot{e} = \dot{\chi} = 0.$
The rest part of the effective action, in fact, coincides with the gauge 
invariant action \cite{Spin} of a spinning particle. One can
interpret the pair $e_0,\chi _0$ in the representation (\ref{d11}) as
a super proper time.

Comparing the path integral representation (\ref{d11}) for
the commutation function  with one \cite{Git} for the Dirac propagator (causal
Green function), one can remark that they are quite similar,  one of
the  differences is in the contour of integration over $s_0$. Namely, the path integral
representation for the Dirac propagator $\tilde{S}^c$ (transformed by
$\gamma^5$) reads
\begin{equation}\label{prop}
\tilde{S}^c =\exp\left(i\tilde{\gamma}^n
\frac{\partial_l}{\partial\theta^n} \right)\int_{\Gamma} \, de_0 \int
d\chi_{0} G(e_0,\chi_0,x_{out}^0,x_{in}^0)\;,
\end{equation}
where the function $G(e_0,\chi_0,x_{out}^0,x_{in}^0)$ has the same
form (\ref{d11a}).  Thus, in case of the commutation function the
c-number component of the super proper time is complex in contrast
with the case of the propagator.

\section{Conclusion}

The results obtained for the commutation function allows one to get
also similar proper time representation for some other singular
functions. For example, it is easy to get for the retarded,
$S^{ret}(x,x')=\Theta(x_0-x'_0)S(x,x')$, and advanced,
$S^{adv}(x,x')=-\Theta(x'_0-x_0)S(x,x')$, functions the following
representations, in which one has to understand $\Theta(0)=1/2$,
 \begin{eqnarray}\label{e36}
&& S^{ret}(x,x')=\Theta(x_0-x'_0)\left( \hat{{\cal P}}_{\nu} \gamma^{\nu} +
m \right)\int_{\Gamma}f(x,x',s)ds\;,\\
&& S^{adv}(x,x')=\Theta(x'_0-x_0)\left( \hat{{\cal P}}_{\nu} \gamma^{\nu} +
m \right)\int_{\Gamma}f(x,x',s)ds\;.
\end{eqnarray}
Combining the Schwinger representation (\ref{e12}) for the causal
Green function  and the
representation (\ref{e19b}) for the commutation function, one can get
proper time representations for positive and negative frequency functions
$S^{\mp}(x,x')$. Namely, let us define them via the Schwinger
representation of the causal Green function, 
\begin{equation}\label{e38}
 S^{\mp}(x,x')=\pm\Theta\left(\pm [x_0-x'_0]\right)S^c(x,x').
\end{equation}
Using the completeness relation
\begin{equation}\label{e39}
 S(x,x')=S^-(x,x')+S^+(x,x')\;,
\end{equation}
we get for any $x,x'$
\begin{equation}\label{e40}
 S^{\mp}(x,x')=\Theta\left(\mp [x_0-x'_0]\right)S(x,x')\pm S^c(x,x').
\end{equation}

In this connection one ought to remark that there is a problem with the
causal Green function definition in case of an arbitrary external
field. From the one hand, there exists the Feynman definition, based on
the definition of the inverse operator to the Dirac equation by means
of the   prescription $m^2 \rightarrow m^2 -i\epsilon$. On the
other hand, in the perturbation theory there appears a field
theoretical definition of the propagator in the form 
\begin{equation}\label{e41a}
S^c(x,x')=i<0|T\psi(x)\bar{\psi}(x')|0>.
\end{equation}
In the absence of the external field or in fields of special form,
which do not violate the vacuum stability (then the operators of the
spinor fields in (\ref{e41a}) have to be taken in the Furry representation), it 
is possible to verify that the Feynman causal Green function and the
propagator (\ref{e12}) coincide. In the same case one can establish that the
former function can be defined via the Schwinger proper time
representation. In external fields, which violate the vacuum stability
(create pairs from the vacuum), the situation is not so clear. In this
case does not exist an unique vacuum for all the time instances. One
has to distinguish the initial $|0,in>$ and final $|0,out >$ vacua
\cite{a2,b100,a3}. In virtue of that, one has also to use different kinds of
propagators in the perturbation theory, 
\begin{eqnarray}
&&S^c(x,x')=i\frac{<0,out|T\psi(x)\bar{\psi}(x')|0,in>}{<0,out|0,in>}\;,\label{e41}\\
&&S^c_{in}(x,x')=i<0,in|T\psi(x)\bar{\psi}(x')|0,in>\;,\label{e42}
\end{eqnarray}
and the positive and negative frequency commutation functions,
\begin{eqnarray}\label{e43}
&&S^-(x,x')=i\frac{<0,out|\psi(x)\bar{\psi}(x')|0,in>}{<0,out|0,in>}\;,\nonumber\\
&&S^+(x,x')=i\frac{<0,out|\bar{\psi}(x')\psi(x)|0,in>}{<0,out|0,in>}\;.
\end{eqnarray}
It was shown that in special cases of external fields, violating the
vacuum stability, the Feynman causal Green function, presented by means
of the Schwinger proper time integral, gives namely the propagator (\ref{e41}),
whereas the propagator (\ref{e42}) demands a modification of the
Schwinger contour in the proper time integration \cite{b101}. At the
present time
 does not exist a proof of the equivalence between
of the Feynman causal Green function and the propagator (\ref{e41}) for any
external fields. Nevertheless,  there is a strong believe 
that they are equivalent. If one excepts such an equivalence, then the positive and
negative frequency commutation functions (\ref{e43}) have the representation (\ref{e40})
in arbitrary external fields.

\noindent 
{\bf Acknowledgement}.

D. Gitman  and S. Gavrilov thank
Brazilian foundations   CNPq  
and

\noindent  FAPESP respectively for support. 
S. Gavrilov thanks also Russian Foundation of Fundamental Research which is
supporting him in part under the Grant No 94-02-03234.
 Both thank Prof. J. Frenkel and Prof. W.F. Wreszinski for 
useful discussions.


\begin{thebibliography}{99}

\bibitem{a1}N.N. Bogoliubov and D.V. Shirkov, {\em Introduction to the
Theory of Quantized Fields} (Wiley-Interscience, New York 1959)
\bibitem{a2}B.S. DeWitt, {\em Dynamical Theory of Groups and Fields}
(Gordon and Breach, New York 1965);  Phys. Rept.C {\bf 19}, 295 (1975)

\bibitem{a3}E.S. Fradkin and D.M. Gitman, Fortschr. Phys. {\bf 29},
381 (1981); E.S. Fradkin, D.M. Gitman and Sh.M. Shvartsman, 
{\em Quantum Electrodynamics with Unstable Vacuum}, (Springer-Verlag, 
Berlin, 1991)

\bibitem{b1a}V. Fock, Phys. Z. Sowjetunion {\bf 12}, 404 (1937)
\bibitem{Schwin} J. Schwinger, Phys. Rev. {\bf 82}, 664 (1951)
\bibitem{b4a}E.S Fradkin and G.A. Vilkovisky, Preprint ITP, University
of Bern (1976); A.O. Barvinsky and G.A. Vilkovisky, Nucl. Phys.  B {\bf
191}, 237 (1981); Phys, Rept. C {\bf 119}, 1 (1985)

\bibitem{Git}E.S. Fradkin and D.M. Gitman, Phys. Rev. {\bf D44} (1991) 3230


\bibitem{Path}E.S. Fradkin, {\em Green's Function Method in Quantized Field
Theory and Quantum Statistics}, Proc. PhIAN {\bf 29} (Nauka, Moscow, 1965) 
[English transl.:Cons. Bureau, N.Y.1967]; M. Henneaux, C. Teitelboim,
Ann. Phys. {\bf 143} (1982) 127; N.V. Borisov and P.P. Kulish, Teor. Math. Fiz. {\bf 51}
(1982) 335; V.Ya. Fainberg and A.V. Marshakov, Nucl. Phys. {\bf B306} (1988)
659; Proc. PhIAN {\bf 201} (1990) 139 (Nauka, Moscow, 1991);
T.M. Aliev, V.Ya. Fainberg, and N.K. Pak, Nucl. Phys. {\bf B429}
(1994) 321; 

\bibitem{Grass}F.A. Berezin, {\em The Method of Second Quantization} (Nauka,
Moscow, 1965); {\em Introduction to Algebra and Analysis with 
Anticommuting Variables} (Moscow State University Press, Moscow, 1983);
{\em Introduction to Superanalysis} (D. Reidel, Dordrecht, 1987);
B. DeWitt, {\em Supermanifolds} (Cambridge Univ. Press, 1985)

\bibitem{ord1}B.S. DeWitt, Rev. Mod. Phys. {\bf 29}, 337 (1951) 

\bibitem{ord2}F.A. Berezin, Uspekhi Fiz. Nauk, {\bf 132}, 497 (1980);
F.A. Berezin and M.A. Shubin, {\em Schr\" odinger Equation} (Moscow
State University, Moscow 1983) 

\bibitem{Spin}F.A. Berezin, M.S. Marinov, Pisma Zh. Eksp. Theor. Fiz. {\bf 21}
 (1975) 678 [JETP Lett. {\bf 21} (1975) 320]; Ann. Phys. (N.Y.) {\bf 104} 
(1977) 336; R. Casalbuoni, Nuovo Cimento {\bf A33} (1976) 115; 389; 
A. Barducci, R. Casalbuoni and L. Lusanna, Nuovo Cimento 
{\bf A35} (1976) 377; L. Brink, S. Deser, B. Zumino, P. di Vechia and P. Howe, Phys. 
Lett. {\bf B64} (1976) 435; L. Brink, P. di Vechia and P. Howe, Nucl. Phys. {\bf B118}  
(1977) 76; A.P. Balachandran, P. Salomonson, B. Skagerstam and J. Winnberg, 
Phys. Rev. {\bf D15}, (1977) 2308               
\bibitem{b100}D.M. Gitman, J. Phys. A {\bf 10}, 2097 (1977)
\bibitem{b101}S.P. Gavrilov, D.M. Gitman and Sh.M. Shvartsman,
Sov. J. Nucl. Phys. (USA) {\bf 29}, 567 (1979); 715 (1979)


\end{thebibliography}
\end{document}